\documentclass[11pt]{article}
\usepackage{jheppub}
\usepackage{amsmath}
\usepackage{amssymb}
\usepackage{braket}
\usepackage{bbold}
\usepackage{pifont}

\newcommand{\be}{\begin{equation}}
\newcommand{\ee}{\end{equation}}
\newcommand{\beqn}{\begin{eqnarray}}
\newcommand{\eeqn}{\end{eqnarray}}
\newcommand{\dd}{\mathrm{d}}
\newcommand{\nn}{\nonumber}
\newcommand{\Tr}{\mathrm{Tr}}

\newcommand{\gmn}{g_{\mu\nu}}

\newcommand{\fmn}{f_{\mu\nu}}

\newcommand{\rmn}{R_{\mu\nu}}

\title{A link between ghost-free bimetric and Eddington-inspired Born-Infeld theory}
\author[1]{Angnis~Schmidt-May,}
\author[2]{Mikael von Strauss}
\affiliation[1]{Department of Physics \& 
        The Oskar Klein Centre,\\
        Stockholm University, AlbaNova University Centre, 
        SE-106 91 Stockholm, Sweden}
        \affiliation[2]{UPMC-CNRS, UMR7095,
 Institut d'Astrophysique de Paris, GReCO,\\
 98bis boulevard Arago, F-75014 Paris, France.}
\emailAdd{angnis.schmidt-may@fysik.su.se}
\emailAdd{strauss@iap.fr}

\abstract{We provide an auxiliary field formulation of the full ghost-free bimetric theory which avoids the explicit presence of a square-root matrix in the action. This description always allows for a branch of solutions where the auxiliary fields can be integrated out to give back the ghost-free theory. For certain parameter regions the two formulations are dynamically equivalent, but in the general case another branch of solutions also exists. We show that this second branch, with certain restrictions on the parameters of the theory, is dynamically equivalent to Eddington-inspired Born-Infeld gravity. This establishes a definite connection between two seemingly unrelated theories of modified gravity.} 


\begin{document} 
\maketitle
\flushbottom

\section{Introduction}

Alternative theories for gravity have been an active field of research ever since the development of general relativity. Within the context of general relativity, modern observations have forced us to introduce a dark matter component as well as dark energy in the form of a cosmological constant. These quantities are inferred gravitationally but lack any evidence of direct detection so far.
Convincing theoretical explanations for the nature of dark matter and a technically natural explanation for the value of the observed cosmological constant are needed for completing our picture of cosmology. 
In recent years, modified gravity theories have received increased attention especially within the context of the dark energy problem. Typically, new degrees of freedom are introduced into the theory in order to replace an unnaturally small cosmological constant term and thereby account for the acceleration of the cosmological expansion. These new degrees of freedom usually come as scalars, vectors or tensors, and they can either be dynamical or merely act as auxiliary fields. 
Besides their application to cosmology and astrophysics, some modifications of general relativity are also employed in the search for a quantum theory of gravity.
For a recent review on the majority of these theories we refer the reader to~\cite{Clifton:2011jh}.

In view of the great variety of different gravitational theories, it is interesting to study possible relations among them in order to narrow down the scope. For instance, $f(R)$ theories, where the Lagrangian density is a function of the Ricci scalar, are classically equivalent to a certain class of scalar-tensor theories~\cite{Chiba:2003ir}.
In this work we will establish a connection (not an equivalence) between two candidates for alternative gravity theories that have received much attention in recent years. On one side we have ghost-free bimetric theory~\cite{Hassan:2011zd}, which is formulated in terms of two rank-two tensors interacting with each other through a nonlinear potential. The kinetic terms in this type of theory are of the standard Einstein-Hilbert form. On the other side there is Eddington-inspired Born-Infeld (EiBI) gravity~\cite{Vollick:2003qp, Vollick:2005gc, Vollick:2006qd, Banados:2010ix} which is formulated in terms of one rank-two tensor and an independent connection field. The action of EiBI theory is nonlinear in the Ricci tensor for the connection and thus its form appears closer to a higher-curvature theory of gravity. On the other hand, as has already been shown in~\cite{Banados:2008fi}, its equations of motion can also be derived from a bimetric-type theory which, however, does not possess the ghost-free structure of~\cite{Hassan:2011zd}. Our aim here is to introduce an action which accommodates both types of theories and thus provides a unified description of EiBI gravity and a subclass of ghost-free bimetric theory. Relations between ghost-free bimetric theory and higher-curvature theories of gravity have already been explored earlier in different setups~\cite{Paulos:2012xe, Hassan:2013pca}.

  \begin{figure}[htbp]
\begin{center}
\includegraphics[width=420pt]{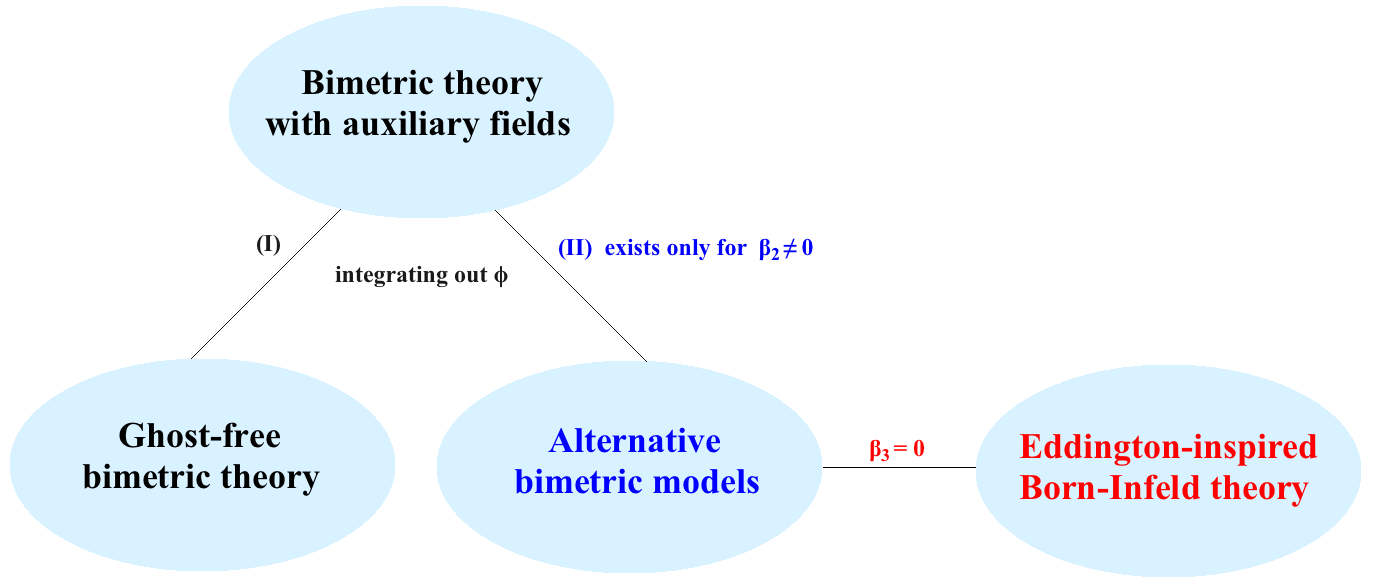}
\label{fig}
\caption{Schematic relation between ghost-free bimetric theory, the auxiliary field formulation and alternative models including EiBI theory. Integrating out the auxiliary fields can be made along two branches when the parameter $\beta_2$ is non-zero. Branch {\bf I} gives back the ghost-free bimetric theory while branch {\bf II} gives a different bimetric theory, a subset of which is dynamically equivalent to EiBI theory.}
\end{center}
\end{figure}

We start from ghost-free bimetric theory including matter and extend its action by introducing a set of auxiliary fields, generalizing an idea first employed in~\cite{Golovnev:2011aa}. In this formulation, the interaction potential does not contain the square-root matrix, which is required for the consistency of bimetric theory and introduces some formal difficulties, see e.g.~\cite{Deffayet:2012zc, Hassan:2014gta}. The new theory with the extended interaction potential is not entirely equivalent to the original bimetric action, but it does contain the latter as part of its solution space.
More precisely, we find that in general the equations of motion for the auxiliary fields possess two branches of solutions: 
While integrating out the fields using the first branch of solutions gives back the original ghost-free bimetric theory, it results in an action with a different structure in the other branch. In a particular parameter subspace where, inter alia, the kinetic term for one of the metrics is removed by setting its ``Planck mass" to zero, the equations of motions in the second branch turn out to be equivalent to those of Eddington-inspired Born-Infeld gravity. The full theory with auxiliary fields therefore gives a unified description of two types of modified gravity theories coupled to matter: It connects EiBI gravity to a version of ghost-free bimetric theory in which one of the metrics has no kinetic term and thus serves as an auxiliary field. These results are illustrated in Figure~\ref{fig}.

In the remainder of the introduction we give brief reviews of ghost-free bimetric as well as EiBI theory. In section~\ref{sec:bimaux} we present the new formulation of bimetric theory involving auxiliary fields in the interaction potential. The connection to EiBI gravity is made in section~\ref{EiBIrel} and our results are summarized and discussed in section~\ref{discussion}.

\subsection{Ghost-free bimetric theory}

Bimetric theories of gravity as alternatives to general relativity have a long history~\cite{Rosen:1940zz, Isham:1971gm}.
Generically, this class of spin-2 theories suffers from an inconsistency known as the Boulware-Deser ghost~\cite{Boulware:1972zf, Boulware:1973my}, which manifests itself as an extra scalar degree of freedom in the spectrum of propagating modes.
An exception to this is a recently developed ghost-free bimetric model which describes consistent nonlinear interactions of a massive with a massless spin-2 field as well as their coupling to matter. It is formulated in terms of two dynamical rank-two tensors $\gmn$ and $\fmn$, which distinguishes it from models for nonlinear massive gravity, in which only one spin-2 field possesses dynamics, while the other is a nondynamical reference metric.
Ghost-free bimetric theory was developed in~\cite{Hassan:2011zd} as a nontrivial generalization of dRGT massive gravity~\cite{deRham:2010ik, deRham:2010kj} and its extension to general reference metrics~\cite{Hassan:2011vm}. All of these theories can be regarded as nonlinear completions of Fierz-Pauli massive gravity~\cite{Fierz:1939ix}, the unique linear model for massive spin-2 fields which avoids the Boulware-Deser ghost instability. The absence of ghost at the nonlinear level was demonstrated in~\cite{Hassan:2011hr, Hassan:2011tf, Hassan:2011ea, Hassan:2012qv} for massive gravity and in~\cite{Hassan:2011zd, Hassan:2011ea} for the bimetric case. The bimetric mass spectrum around maximally symmetric solutions has been studied in~\cite{Hassan:2012wr, Schmidt-May:2014xla} and there exists a large amount of literature on the cosmology of the theory. For a list of references we refer the reader to the recent review~\cite{deRham:2014zqa}.

Let us briefly summarize the relevant technical details of ghost-free bimetric theory. The two rank-two tensors $\gmn$ and $\fmn$ possess Einstein-Hilbert kinetic terms and interact with each other through a nonlinear potential.
More explicitly, the bimetric action is of the form~\cite{Hassan:2011zd},
\beqn\label{biact}
S_\mathrm{bi}[g,f]&=&m_g^2\int\dd^4x\sqrt{g}\,\big(R(g)-2\Lambda_g\big)-2m^2m_g^2\int\dd^4x~V(g,f)\nn\\
&~&+~m_f^2\int\dd^4x\sqrt{f}\,\big(R(f)-2\Lambda_f\big)+\int\dd^4x\sqrt{f}~\mathcal{L}_\mathrm{matter}(f,\psi)\,,
\eeqn
where $m_g$,$m_f$ are generalized Planck masses parameterizing the interaction strengths of the respective metrics and $m$ is another mass scale which sets the mass of the massive spin-2 mode.
The structure of the interaction potential $V(g,f)$ is fixed by demanding the absence of the Boulware-Deser ghost instability. It contains three free parameters $\beta_n$ and reads~\cite{deRham:2010kj, Hassan:2011vm},
\beqn\label{pot}
V(g,f)=\sqrt{g}\sum_{n=1}^3\beta_n e_n\big(\sqrt{g^{-1}f}\,\big)\,,
\eeqn
in which $\sqrt{g^{-1}f}$ is a square-root matrix defined\footnote{Note that the definition $\sqrt{\mathbb{X}}\sqrt{\mathbb{X}}=\mathbb{X}$ for a matrix square root also implies $\sqrt{\mathbb{X}^2}=\mathbb{X}$.} via $(\sqrt{g^{-1}f}\,)^2=g^{-1}f$ and $e_n(\mathbb{X})$ are the elementary symmetric polynomials of any matrix $\mathbb{X}$. Their explicit expressions are,
\beqn\label{endef}
e_1(\mathbb{X})=\Tr[\mathbb{X}]\,,\qquad 
e_2(\mathbb{X})=\tfrac{1}{2}\big((\Tr[\mathbb{X}])^2-\Tr[\mathbb{X}^2]\big)\,,\nn\\
e_3(\mathbb{X})=\tfrac{1}{6}\big((\Tr[\mathbb{X}])^3-3\Tr[\mathbb{X}^2]\Tr[\mathbb{X}]+2\Tr[\mathbb{X}^3]\big)\,.
\eeqn
It follows from the identity $e_n(\mathbb{X}^{-1})=\det(\mathbb{X})\,e_{4-n}(\mathbb{X})$ obeyed by the elementary symmetric polynomials that the potential (\ref{pot}) may also be written as,
\beqn
V(g,f)=\sqrt{f}\sum_{n=1}^3\beta_{4-n} e_n\big(\sqrt{f^{-1}g}\,\big)\,.
\eeqn
The structure of the action (\ref{biact}) is therefore symmetric under the interchange of the two metrics, except for the matter coupling. Due to the re-appearance of the Boulware-Deser ghost, it is not possible to add a coupling between $\gmn$ and the same matter fields $\psi$ that enter the matter Lagrangian $\mathcal{L}_\mathrm{matter}(f,\psi)$ for $\fmn$~\cite{Yamashita:2014fga, deRham:2014naa}. One can therefore only couple matter to one of the metrics, which we have chosen to be $\fmn$ here for later purposes.

The appearance of the square-root matrix $\sqrt{g^{-1}f}$ can sometimes lead to technical difficulties and require rather large computational effort, which becomes evident for instance in the ghost proof~\cite{Hassan:2011hr, Hassan:2011tf, Hassan:2011ea, Hassan:2011zd} or in the study of perturbation theory around general solutions~\cite{Bernard:2014bfa}.

\subsection{Eddington-inspired Born-Infeld theory}

In 1924 Eddington proposed a theory for gravity not in terms of a metric, but based solely on a connection field~\cite{Eddington}. Even though Eddington's gravity is equivalent to general relativity in vacuum, it is incomplete since it does not account for interactions of gravity with matter.
One possibility to overcome this drawback and introduce matter couplings is to give up on the equivalence to Einstein's theory and allow the action to contain a metric besides the connection field.\footnote{For attempts to include matter couplings into Eddington's original theory see, for instance,~\cite{Ferraris:1981hr, Poplawski:2006zr}.}
Such an extension of Eddington's theory including matter fields has been developed in~\cite{Vollick:2003qp, Vollick:2005gc, Vollick:2006qd, Banados:2010ix}, resulting in an alternative theory for gravity with Born-Infeld structure~\cite{Born:1934gh}. This Eddington-inspired Born-Infeld (EiBI) theory has received much attention in recent years and its implications for cosmology and astrophysics have been the subject of extensive study, for example in~\cite{EscamillaRivera:2012vz, Cho:2012vg, Scargill:2012kg, Jimenez:2014fla}.
In particular, it has been shown that the theory avoids singularities that typically arise in general relativity, such as black hole singularities or the initial cosmic singularity~\cite{Banados:2010ix, Pani:2011mg, Avelino:2012ue}.
On the other hand, it seems to exhibit unobserved astrophysical singularities~\cite{Pani:2012qd}, which unfortunately cast serious doubts on the phenomenological viability of the theory in its present form.

The EiBI action for a metric $\fmn$ and an independent connection field $\Gamma^\mu_{\rho\sigma}$ is of the following form~\cite{Vollick:2003qp, Vollick:2005gc, Vollick:2006qd, Banados:2010ix},
\beqn\label{eibiactor}
S_\mathrm{EiBI}[f, \Gamma]&=&\frac{2}{\kappa^2}\int\dd^4x\Bigg(\sqrt{\det\big(\fmn+\kappa \rmn(\Gamma)\big)}-\lambda \sqrt{f}\Bigg)+\int\dd^4x\sqrt{f}~\mathcal{L}_\mathrm{matter}(f,\psi)\,.
\eeqn
Here, $\kappa$ is a parameter of mass dimension $-2$, $\lambda$ is a dimensionless constant and $\rmn(\Gamma)$ is the symmetric part of the Ricci tensor built from the connection.\footnote{Note that $\rmn$ in general does not have to be symmetric but that only its symmetric part appears in the standard formulation of EiBI theory, hence we take it to be symmetric in the following.} The parameters in the action are chosen such that for small values of $\kappa R$, one recovers the familiar Einstein-Hilbert action with cosmological constant $\tilde\Lambda=\kappa^{-1}(\lambda-1)$.
More complicated versions of the model could be considered, in which the connection field also enters the matter Lagrangian $\mathcal{L}_\mathrm{matter}$ but here we choose to follow the simplest route and couple the matter fields $\psi$ only to the metric.
The equations of motion for $\fmn$ following from the above action are then given by,
\beqn\label{firsteq}
0=\sqrt{\det\big(\fmn+\kappa \rmn(\Gamma)\big)}\big((f+\kappa R)^{-1}\big)^{\mu\nu}-\lambda \sqrt{f}~f^{\mu\nu}+\kappa^2 \sqrt{f}~T^{\mu\nu}\,,
\eeqn
where $\sqrt{f}\,T^{\mu\nu}=\frac{\partial(\sqrt{f}\,\mathcal{L}_\mathrm{matter})}{\partial f_{\mu\nu}}$ is the stress-energy tensor for the matter source. 
Next, we would also like to compute the equations of motions associated to the connection field.
It turns out that the simplest way to proceed is to introduce an auxiliary metric~$\gmn$ which is compatible with the connection $\Gamma^\mu_{\rho\sigma}$. Its equations of motion derived from~(\ref{eibiactor}) then take the simple form~\cite{Banados:2010ix, Scargill:2012kg},
\beqn\label{secondeq}
0=\kappa\rmn(\Gamma)-\gmn+\fmn\,,
\eeqn
where now $\Gamma^\mu_{\rho\sigma}=\frac{1}{2}g^{\mu\lambda}(\partial_\rho g_{\sigma\lambda}+\partial_\sigma g_{\rho\lambda}-\partial_\lambda g_{\rho\sigma})$. 
Combining (\ref{firsteq}) and (\ref{secondeq}), one finally arrives at the following equivalent system of equations,
\begin{subequations}\label{eibieqs}
\beqn
0&=&\kappa\rmn(\Gamma)-\gmn+\fmn\,,\label{eibieqs1}\\
0&=&\sqrt{g}~g^{\mu\nu}-\lambda\sqrt{f}~ f^{\mu\nu}+\kappa^2\sqrt{f}~T^{\mu\nu}\,.\label{eibieqs2}
\eeqn
\end{subequations}
In vacuum, where $T^{\mu\nu}=0$, equation~(\ref{eibieqs2}) can be solved for $\fmn$ to give $ f_{\mu\nu}=\lambda\sqrt{\det(g^{-1}f)}\,g_{\mu\nu}$. Plugging this into (\ref{eibieqs1}), taking the divergence and using the Bianchi identity shows that $\sqrt{\det(g^{-1}f)}=$\,const.~and hence,
\beqn
R_{\mu\nu}=\Lambda\gmn\,,\quad f_{\mu\nu}=\big(1-\kappa\Lambda\big)g_{\mu\nu}\,,\quad \text{where}~~\Lambda=\frac{1}{\kappa}\Big(1-\lambda\sqrt{\det(g^{-1}f)}\Big)=\text{const.}
\eeqn
Both metrics thus solve Einstein's vacuum equations in the presence of a cosmological constant. Hence, in the absence of matter, EiBI theory is classically equivalent to general relativity. In particular, it contains only the two propagating degrees of freedom of a massless graviton. Differences between the two theories occur only when matter is present.

\section{Auxiliary field formulation of bimetric theory}\label{sec:bimaux}

In the following, we write down an extension of ghost-free bimetric theory which contains a set of nondynamical auxiliary fields $\Phi^\mu_{~\nu}$ in addition to the two metrics $\gmn$ and $\fmn$. We begin by presenting the action~$S[g,f,\Phi]$ with general parameters and discuss the different branches of solutions to the $\Phi^\mu_{~\nu}$ equations of motion. We demonstrate that there always exist a ``natural" branch on which the action reduces to the ghost-free bimetric theory~(\ref{biact}) for the two remaining metrics. Other solutions may exist but are more difficult to interpret in the general case. We therefore proceed by focussing on special parameter choices for which the additional solutions are either absent or for which the auxiliary fields can be integrated out in a straightforward way. We end this section with a brief overview of the logic used to obtain the auxiliary field extension and discuss some ambiguities in its formulation.

\subsection{Full bimetric action with auxiliary fields}\label{sec:auxfull}

The action that we will consider in what follows has the following structure,
\beqn\label{genauxact}
S[g,f,\Phi]&=&m_g^2\int\dd^4x\sqrt{g}\,\big(R(g)-2\Lambda_g\big)-2m^2m_g^2\int\dd^4x~V(g,f,\Phi)\nn\\
&~&+~m_f^2\int\dd^4x\sqrt{f}\,\big(R(f)-2\Lambda_f\big)+\int\dd^4x\sqrt{f}~\mathcal{L}_\mathrm{matter}(f, \psi)\,,
\eeqn
in which the interaction potential containing the two metric $\gmn$ and $\fmn$ as well as the auxiliary fields~$\Phi^\mu_{~\nu}$ consists of three terms,
\beqn
V(g,f,\Phi)=V_1+V_2+V_3\,.
\eeqn
Its components are parametrized by three free parameters $\beta_1$, $\beta_2$ and $\beta_3$ and read,
\begin{subequations}\label{potdef}
\beqn
V_1&=&\frac{\beta_1}{2}\sqrt{g}~\Tr\big[ \Phi+\Phi^{-1}g^{-1}f  \big]\,,\label{v1}\\
V_2&=& \frac{\beta_2}{8}\sqrt{g}\left(\Big(\Tr\big[ \Phi+\Phi^{-1}g^{-1}f  \big]\Big)^2-4\,\Tr\big[g^{-1}f\big]\right)\,,\label{v2}\\
V_3&=&\frac{\beta_3}{2}\sqrt{f}~\Tr\big[ \Phi^{-1}+\Phi f^{-1}g  \big]\,.\label{v3}
\eeqn
\end{subequations}
The structure of these interactions has been chosen such that it can reproduce the ghost-free bimetric terms, but this criterion does not uniquely fix the form of~(\ref{potdef}). We comment on this ambiguity and the general motivation behind the structure of the above interactions in more detail in section~\ref{sec:auxamb}.

The complete equations of motions for the auxiliary fields $\Phi^\nu_{~\mu}$ following from the above action are given by,
\beqn\label{phieqott}
0=\sqrt{g}\Big(\beta_1&+&\tfrac1{2}\beta_2\Tr\Big[ \Phi+\Phi^{-1}g^{-1}f  \Big]\Big)\Big( \delta^\mu_{~\nu}- (\Phi^{-1})^\mu_{~\rho} g^{\rho\sigma}f_{\sigma\kappa} (\Phi^{-1})^\kappa_{~\nu}\Big)\nn\\
&~&\hspace{150pt}+ ~
 \beta_3\sqrt{f}\Big(  f^{\mu\sigma}g_{\sigma\nu}- (\Phi^{-2})^\mu_{~\nu} \Big)
 \,.
\eeqn
Our aim is to find solutions to these equations which allow us to integrate out the auxiliary fields from the action.
To this end, let us multiply the equations with $\Phi^\rho_{~\mu}\Phi^\nu_{~\sigma}$, after which they are of the equivalent form,\footnote{It is implicit in our construction that we assume the matrix of auxiliary fields to be invertible.}
\beqn\label{int2}
\Phi S^{-1} M S^{-1}\Phi=M\,.
\eeqn
Here we have switched to matrix notation and used the definitions $S^\mu_{~\nu}\equiv(\sqrt{g^{-1}f}\,)^\mu_{~\nu}$ as well as,
\beqn\label{defm}
M^\rho_{~\sigma}\equiv\beta_3\sqrt{f}~\delta^\rho_{~\sigma}+\Big(\beta_1+\tfrac1{2}\beta_2\Tr\Big[ \Phi+\Phi^{-1}g^{-1}f  \Big]\Big)\sqrt{g}~g^{\rho\alpha}f_{\alpha\sigma}\,.
\eeqn
Since this expression for the matrix $M$ is dependent on the auxiliary fields, the general equations of motion do not have a unique solution for~$\Phi$, as we will see in the following. Instead, the solutions split up into two branches which are discussed separately below. For an illustration of the results we refer to Figure~\ref{fig}.

\subsubsection*{Branch {\bf I}}
One solution to (\ref{int2}) is obtained assuming that the matrix $M$ is invertible. In this case, we multiply the equation with $S^{-1}MS^{-1}$ from the right and obtain,
\beqn
\Phi S^{-1}MS^{-1}\Phi S^{-1}MS^{-1}= MS^{-1}MS^{-1}\,.
\eeqn
Note that both sides of this equation are perfect squares.
Taking the square root of the equation and multiplying by the inverse of $S^{-1}MS^{-1}$ (which exists by assumption in this branch) from the right gives,
\beqn
{\bf{(I)}}~~\Phi^\mu_{~\nu}=\pm(\sqrt{g^{-1}f}\,)^\mu_{~\nu}\,.
\eeqn
Plugging this expression back into the interaction terms in (\ref{potdef}), we obtain,
\beqn
V(g,f)=\sqrt{g}\left(\pm \beta_1 e_1\big(\sqrt{g^{-1}f}\,\big)+ \beta_2 e_2\big(\sqrt{g^{-1}f}\,\big)\pm \beta_3 e_3\big(\sqrt{g^{-1}f}\,\big).\right)\,,
\eeqn
where for the $\beta_3$-term we have used the identity $e_1(\mathbb{X}^{-1})=\det(\mathbb{X})e_3(\mathbb{X})$ for a matrix $\mathbb{X}$. This shows that integrating out the auxiliary fields from (\ref{genauxact}) on this solution branch results in the full ghost-free bimetric action~(\ref{biact}). Hence, when restricting to this branch of solutions for~$\Phi$, we have an auxiliary field description for ghost-free bimetric theory without the appearance of any square-root matrix in the action. 

\subsubsection*{Branch {\bf II}}
On the other branch, the matrix $M$ is not invertible and thus we must have,
\beqn\label{deteq}
{\bf{(II)}}~~0=\det M=\det\left[\beta_3\sqrt{f}~\mathbb{1}+\Big(\beta_1+\tfrac1{2}\beta_2\Tr\Big[ \Phi+\Phi^{-1}g^{-1}f  \Big]\Big)\sqrt{g}~g^{-1}f\right]\,.
\eeqn
This equation is a fourth-order polynomial in the trace combination $\tau\equiv\Tr\left[ \Phi+\Phi^{-1}g^{-1}f  \right]$ and hence implies,
\beqn\label{trone}
{\bf{(II)}}~~\tau=F(g^{-1}f)\,,
\eeqn
where $F(g^{-1}f)$ is a complicated scalar function of $g^{-1}f$ corresponding to any of the four roots of~(\ref{deteq}). It is therefore possible to extract a solution for~$\tau$ from the above scalar equation but, since the $\beta_3$-term contains a different trace combination than~$\tau$, we need a full solution for~$\Phi^\mu_{~\nu}$ in order to completely integrate out the fields. 
Deriving a covariant expression for~$\Phi^\mu_{~\nu}$ in terms of $\gmn$ and $\fmn$ from equation~(\ref{int2}) seems difficult, however.
Even though we may easily construct a solution for $\Phi^\mu_{~\nu}$ which after tracing is consistent with~\eqref{trone}, this only means that~$M$ is not invertible and we still have to make sure that the full equations of motion~\eqref{int2} are satisfied. It seems that these equations will eventually constrain the configuration of solutions for the metrics $\gmn$ and $\fmn$ instead of determining merely the auxiliary fields.
For general parameters, it is therefore not obvious if it is even possible to integrate out the auxiliary fields and obtain an interesting covariant bimetric action for $\gmn$ and~$\fmn$ in this branch. 

On the other hand, note that the interaction terms (\ref{v1}) and (\ref{v2}) are functions of $\tau$ only, whereas only (\ref{v3}) contains a different type of trace. Thus we can actually integrate out the auxiliary fields explicitly on branch~{\bf II} if $\beta_3=0$. This case, which turns out to yield the connection to EiBI theory, will be discussed in more detail in section~\ref{btwo}.  Before studying the details of the $\beta_3=0$ case, we briefly discuss two other parameter choices, for which branch {\bf II} does not exist and hence the auxiliary action is classically equivalent to ghost-free bimetric theory.

\subsection{The $\beta_1$-model}\label{b1model}

In~\cite{Golovnev:2011aa} auxiliary fields of the above type were introduced to reformulate the potential of a particular ghost-free bimetric model without the appearance of the square-root matrix $\sqrt{g^{-1}f}$ in the action. This model corresponds to (\ref{potdef}) restricted to $\beta_2=\beta_3=0$ and we will briefly review it here.

If only $\beta_1$ is nonzero, the equations of motion~(\ref{phieqott}) for $\Phi^\nu_{~\mu}$ reduce to,
\beqn\label{boeq}
 \delta^\mu_{~\nu}- (\Phi^{-1})^\mu_{~\rho} g^{\rho\sigma}f_{\sigma\kappa} (\Phi^{-1})^\kappa_{~\nu} =0\,.
\eeqn
The matrix $M$ in (\ref{defm}) is now simply $M=\beta_1\sqrt{g}\,\,g^{-1}f$ which is invertible by the default assumption of invertibility of $g$ and $f$. As a consequence, branch {\bf I} gives the unique solution,
\beqn
\Phi^\mu_{~\nu}=\pm\big(\sqrt{g^{-1}f}\,\big)^\mu_{~\nu}\,.
\eeqn
Plugging this solution for the auxiliary fields back into the action results in an interaction potential for $\gmn$ and $\fmn$ alone,
\beqn
V_1=\pm\beta_1\sqrt{g}\,\Tr\Big[ \sqrt{g^{-1}f}  \Big]=\pm\beta_1\sqrt{g}\,e_1\big(\sqrt{g^{-1}f}\,\big)\,,
\eeqn
which is exactly the $\beta_1$-term of the ghost-free bimetric potential \eqref{pot}. The auxiliary field formulation is therefore classically equivalent to the corresponding bimetric model. The authors of \cite{Golovnev:2011aa, Hassan:2012qv} made use of this fact to perform an ADM analyses of the $\beta_1$-model and confirm the absence of the Boulware-Deser ghost. Moreover, \cite{Buchbinder:2012wb} invoked this auxiliary field formulation to compute quantum effects in this model.

\subsection{Including the $\beta_3$-term}\label{bthree}

We now add the term $V_3$ in (\ref{v3}) to the potential, in order to also recover the $\beta_3$-term of ghost-free bimetric theory. This term will not be needed for making the connection to EiBI theory, but it constitutes the simplest extension of the simple case discussed in the previous section.
Let us consider the potential (\ref{potdef}) with $\beta_2=0$, for which the equations of motion for $\Phi^\nu_{~\mu}$ given in~(\ref{int2}) reduce to,
\beqn\label{int1}
\Phi^\rho_{~\mu}(S^{-1})^{\mu}_{~\lambda}M^\lambda_{~\kappa}(S^{-1})^{\kappa}_{~\nu}\Phi^\nu_{~\sigma}=M^\rho_{~\sigma}\,,
\eeqn
with,
\beqn\label{mdef1}
M^\rho_{~\sigma}=\beta_3\sqrt{f}~\delta^\rho_{~\sigma}+\beta_1\sqrt{g}~g^{\rho\alpha}f_{\alpha\sigma}\,,\qquad
S^\mu_{~\nu}=(\sqrt{g^{-1}f}\,)^\mu_{~\nu}\,.
\eeqn
As in the $\beta_1$-model, the matrix $M$ is independent of $\Phi$. This allows us to assume the existence of its inverse $M^{-1}$, because otherwise the $\Phi$-equations would constrain $g$ and $f$ instead of yielding solutions for $\Phi$. For generic metrics $g$ and $f$, the inverse certainly exists.\footnote{More precisely, the condition $\det(M)=0$ can be viewed as a 4th order polynomial in $\beta_1/\beta_3$ which can be solved in terms of the eigenvalues of $S=\sqrt{g^{-1}f}$. This situation can only occur for a very finely tuned relation between $\beta_1$ and $\beta_3$ together with a particular configuration of $g$ and $f$ and is therefore hardly generic.
For example, the simplest exception $M=0$ directly implies $\beta_1/\beta_3=-e_1^2(S)/16$ and $S=\pm\sqrt{-\beta_1/\beta_3}\,\mathbb{1}$.
}
Following the subsequent steps outlined in section~\ref{sec:auxfull} we thus again find that branch~{\bf I} with,
\beqn\label{phisol13}
\Phi^\mu_{~\nu}=\pm(\sqrt{g^{-1}f}\,)^\mu_{~\nu}\,,
\eeqn
is the unique solution to the equations. Plugging this back into the potential returns the ghost-free bimetric interactions with $\beta_2=0$,
\beqn
V_1+V_3&=&\pm\beta_1\sqrt{g}\,\Tr\Big[ \sqrt{g^{-1}f}  \Big]\pm \beta_3\sqrt{f}\,\Tr\Big[ \sqrt{f^{-1}g}  \Big]\nn\\
&=&\pm\sqrt{g}\,\Big(\beta_1e_1\big(\sqrt{g^{-1}f}\,\big)+\beta_3e_3\big(\sqrt{g^{-1}f}\,\big)\Big)\,,
\eeqn
where we have used the identity $\sqrt{f}\,e_1(S^{-1})=\sqrt{g}\,e_3(S)$. The overall sign depends on which sign is chosen for the solution in~(\ref{phisol13}).
This shows that the auxiliary action $S[g,f,\Phi]$ of the form (\ref{genauxact}) with $\beta_2=0$ is classically equivalent to the ghost-free bimetric action~(\ref{biact}) with $\beta_2=0$.
We have thus extended the auxiliary field formulation to this class of models, which enables us to formulate them without having to introduce the square-root matrix into the action.

\subsection{Including the $\beta_2$-term}\label{btwo}

Now we turn to the auxiliary field formulation that can also generate the ghost-free bimetric $\beta_2$-term. In order to be able to integrate out the auxiliary fields explicitly in both branches, we set $\beta_3=0$, keeping only $\beta_1$ and $\beta_2$ nonzero in~(\ref{potdef}). 
The equations for $\Phi^\nu_{~\mu}$ in this model read,
\beqn\label{bteq}
\Big(2\beta_1+\beta_2\Tr\Big[ \Phi+\Phi^{-1}g^{-1}f  \Big]\Big)\Big( \delta^\mu_{~\nu}- (\Phi^{-1})^\mu_{~\rho} g^{\rho\sigma}f_{\sigma\kappa} (\Phi^{-1})^\kappa_{~\nu} \Big)
=0\,.
\eeqn
Note that the second bracket in this expression resembles the equation~(\ref{boeq}) of the $\beta_1$-model.
But in this case there exist two branches of solutions, corresponding to the vanishing of either the rightmost or the leftmost bracket respectively,
\beqn\label{brbb}
{\text{\bf(I)}}~~\Phi^\mu_{~\nu}=\pm\big(\sqrt{g^{-1}f}\,\big)^\mu_{~\nu}\,,\qquad \text{and}\qquad
{\text{\bf(II)}}~~\Tr\Big[ \Phi+\Phi^{-1}g^{-1}f  \Big]=-\frac{2\beta_1}{\beta_2}\,.
\eeqn
The equation of branch {\bf II} is of course equivalent to the general condition $\det M=0$ obtained in \eqref{deteq}, since the matrix~$M$ in~(\ref{int2}) now reads,
\beqn
M^\rho_{~\sigma}=\Big(\beta_1+\tfrac1{2}\beta_2\Tr\Big[ \Phi+\Phi^{-1}g^{-1}f  \Big]\Big)\sqrt{g}~g^{\rho\alpha}f_{\alpha\sigma}\,.
\eeqn
Branch {\bf I}, which is obtained from setting the rightmost bracket in (\ref{bteq}) to zero, is similar to the $\beta_2=0$ case. 
Plugging this solution back into the potential, we get,
\beqn
V^{\mathrm{(I)}}_1+V^{\mathrm{(I)}}_2&=&\pm\beta_1\sqrt{g}\,\Tr\Big[ \sqrt{g^{-1}f}  \Big]+\beta_2\frac{\sqrt{g}}{2}\Bigg(\Big(\,\Tr\Big[ \sqrt{g^{-1}f}  \Big]\Big)^2-\Tr\Big[g^{-1}f\Big]\Bigg)\nn\\
&=&\sqrt{g}\Big(\pm\beta_1e_1\big(\sqrt{g^{-1}f}\,\big)+\beta_2e_2\big(\sqrt{g^{-1}f}\,\big)\Big)\,,
\eeqn
which is exactly the potential of ghost-free bimetric theory with $\beta_3=0$. The sign in front of the $\beta_1$-term again depends on which sign for $\Phi$ is chosen in branch {\bf I} of (\ref{brbb}).

On the other hand, for the branch {\bf II} solution, we can now integrate out the auxiliary fields explicitly since for $\beta_3=0$ they appear in the action only through the particular trace combination~$\tau=\Tr\big[ \Phi+\Phi^{-1}g^{-1}f  \big]$. 
The value for $\tau$ in branch {\bf II} can be directly read off from (\ref{brbb}) and the potential becomes,
\beqn\label{btpot}
V^{\mathrm{(II)}}_1+V^{\mathrm{(II)}}_2=-\frac{\beta_1^2}{2\beta_2}-\frac{\sqrt{g}}{2}\Tr\Big[g^{-1}f\Big]\,,
\eeqn
which does not have the structure of the ghost-free bimetric theory. The auxiliary field formulation is therefore not equivalent to the corresponding ghost-free bimetric model when $\beta_2\neq0$, but rather contains it as part of its solution space. In addition, it contains solutions for $\gmn$ and $\fmn$ which follow from a bimetric action with interaction potential~(\ref{btpot}). This part of the solution space has a very concrete connection to the EiBI theory, as we will discuss further in section~\ref{EiBIrel}.

\subsection{Ambiguities in the auxiliary formulation}\label{sec:auxamb}

We recall that the general motivation behind the auxiliary field formulation is to remove any explicit appearances of the square-root matrix $S=\sqrt{g^{-1}f}$. To this end, the author of~\cite{Golovnev:2011aa} first considered the model with only $V_1=\beta_1\sqrt{g}\,e_1(S)$ in the interaction potential and made the replacement $S\rightarrow A/2$ where $A=\Phi+\Phi^{-1}\,g^{-1}f$, as reviewed in section \ref{b1model}. Already here an initial ambiguity\footnote{Other ambiguities may also arise. For example, in a footnote of~\cite{Golovnev:2011aa} the author suggests a different generalization in which the $\beta_2$-term has the form, $V_2\sim\Tr[\Phi]\Tr[\Phi^{-1}S^2]-\Tr[S^2]$. In this case, if $\beta_1=\beta_3=0$, the equations imply $\Phi=\pm \Omega\, S$ with a completely undetermined scalar function $\Omega$, and the auxiliary formulation is dynamically equivalent to the corresponding ghost-free model. For general parameters, there may again exist other solutions.  This type of auxiliary potential cannot give a connection to EiBI theory and we do therefore not explore this alternative option in more detail here.} arises because of the fact that the elementary symmetric polynomials satisfies the identity $e_n(\mathbb{X})=\det(\mathbb{X})\,e_{4-n}(\mathbb{X}^{-1})$ for any matrix $\mathbb{X}$, which translates to,
\beqn\label{ensym}
\sqrt{g}~e_n(S)=\sqrt{f}~e_{4-n}(S^{-1})\,.
\eeqn
This means that we could instead consider the equivalent expression $V_1=\beta_1\sqrt{f}\,e_3(S^{-1})$ with the form of $e_3$ given in \eqref{endef}. Here we could replace any single powers of $S^{-1}$ through $S^{-1}\rightarrow B/2$ where $B=\Phi^{-1}+\Phi f^{-1}g$ (meaning that in the term with $\Tr[S^{-3}]$ we replace only one power of $S^{-1}$) and get an alternative auxiliary formulation,
\beqn
V_1=\tfrac{\beta_1\sqrt{f}}{6}\left(\tfrac{1}{8}\Big(\Tr[ B ]\Big)^3-\tfrac{3}{2}\,\Tr[f^{-1}g]\Tr[ B ]+\Tr[ Bf^{-1}g  ]\right)\,.
\eeqn
It is easy to see that $\Phi^\mu_{~\nu}=\pm S^\mu_{~\nu}$ solves the equations for the auxiliary fields also here and, on this solution branch, the potential for the metrics $\gmn$ and $\fmn$ assumes the ghost-free structure $V^{(I)}_1=\beta_1\sqrt{f}\,e_3(S^{-1})=\beta_1\sqrt{g}\,e_1(S)$. However, since $2A^{-1}\neq B/2$ the two ways of introducing auxiliary fields are not equivalent. In fact, when using this second option there exists also another branch of solutions, for which it is not obvious how to fully integrate out the auxiliary fields. Therefore the first option of replacing $S$ by $A/2$ to introduce the auxiliary fields seems preferred. A completely analogous discussion can be made for the model with only $V_3=\beta_3\sqrt{g}\,e_3(S)=\beta_3\sqrt{f}\,e_1(S^{-1})$ but here the option of replacing $S^{-1}$ by $B/2$ to introduce the auxiliary fields turns out to be preferred for exactly the same reasoning as above. 

For the model with only $V_2=\beta_2\sqrt{g}\,e_2(S)=\beta_2\sqrt{f}\,e_2(S^{-1})$ there is no obvious way of discriminating between the two options, it only comes down to an interchange of the roles of the two metrics $\gmn$ and $\fmn$ in the final potential.  
As we saw in the previous section, even when we added a nonzero $\beta_1$-term we could integrate out the auxiliary fields fully in branch~{\bf II} with our choice of the $\beta_2$-term (the $A/2$ replacement). Had we instead chosen the second option for the $\beta_2$-term (the $B/2$ replacement) we would reach the same conclusion but should now add it to a nonzero $\beta_3$- instead of a $\beta_1$-term. Then we would have obtained a potential of exactly the same form as \eqref{btpot} but with $g$ and $f$ interchanged as well as $\beta_3$ replacing $\beta_1$. On the other hand, if both type of terms had been introduced at the same time, we would no longer be able to integrate out the auxiliary fields fully in the second branch. Therefore it is preferrable to use only one of the options and not both simultaneously.

In principle, one could also consider including all types of terms together in the action and recover the ghost-free structure in one solution branch, but in this case it would again not be possible to fully integrate out the auxiliary fields and obtain a covariant bimetric action in the second branch.    

In summary, the reasoning for singling out the formulation in (\ref{potdef}) is as follows: For the $\beta_1$- and $\beta_3$-terms our way of introducing the auxiliary fields is the preferred one because it is equivalent to the ghost-free theory when $\beta_2=0$. In the $\beta_2$-term, in order to maintain the possibility to eliminate the auxiliary fields from the action whenever $\beta_3=0$ (or $\beta_1=0$), we only keep one of the two possible terms. Then the ambiguity merely corresponds to an interchange of the two metrics $\gmn$ and $\fmn$ and we have picked one of the two options here without any direct loss of generality.

\section{Connection to Eddington-inspired Born-Infeld theory}\label{EiBIrel}

In this section we show how the branch {\bf II} solutions for the auxiliary fields in the model of section~\ref{btwo} with vanishing $\beta_3$ is equivalent to Eddington-inspired Born-Infeld (EiBI) theory. To this end, we first briefly review the bimetric formulation of the latter.

\subsection{Bimetric formulation of EiBI theory}

It has already been pointed out in~\cite{Banados:2008fi}, that the equations of motion~(\ref{eibieqs}) for EiBI theory also follow from an auxiliary action which is of bimetric type,
\beqn\label{auxeibi}
S'_\mathrm{bi}[g,f, \Gamma]&=&\int\dd^4x \Bigg(\frac{\sqrt{g}}{\kappa}\Big(g^{\mu\nu}\rmn(\Gamma)-\frac{2}{\kappa}+\frac{1}{\kappa}\Tr\big[g^{-1}f\big]\Big)\nn\\
&~&\hspace{137pt}
+\sqrt{f}\left(\mathcal{L}_\mathrm{matter}(f,\psi)-\frac{2\lambda}{\kappa^2}\right)\Bigg)\,.
\eeqn 
Here, $\Gamma$ is an independent connection field whose equations of motion have the solution $\Gamma^\mu_{\rho\sigma}=\frac{1}{2}g^{\mu\lambda}(\partial_\rho g_{\sigma\lambda}+\partial_\sigma g_{\rho\lambda}-\partial_\lambda g_{\rho\sigma})$. Thus, integrating out the connection field results in a bimetric action for the two tensor fields $\gmn$ and $\fmn$.
On the other hand, integrating out $\gmn$ in~(\ref{auxeibi}) gives back the original EiBI action $S_\mathrm{EiBI}[f,\Gamma]$ in~(\ref{eibiactor}). In order to see this, consider the equations of motion following from varying (\ref{auxeibi}) with respect to $g^{\mu\nu}$,
\beqn\label{eomgei}
0=R_{\mu\nu}(\Gamma)-\tfrac{1}{2}g^{\rho\sigma}R_{\rho\sigma}(\Gamma)\gmn+\frac{1}{\kappa}\gmn+\frac{1}{\kappa}\Big(\fmn-\tfrac{1}{2}\gmn\Tr\big[g^{-1}f\big] \Big)\,.
\eeqn
By subtracting half of the traced equation, one can obtain the solution for $\gmn$ as,
\beqn\label{geq}
\gmn=\kappa \rmn(\Gamma)+\fmn\,,
\eeqn
which is of the same form as equation~(\ref{eibieqs1}). Plugging the solution for $\gmn$ back into (\ref{auxeibi}) gives back the EiBI action in~~(\ref{eibiactor}).  Note that carrying out this procedure is possible only because the interaction potential has the simple structure $\sqrt{g}\,\Tr\big[g^{-1}f\big]$ and equation~(\ref{eomgei}) can easily be solved for $\gmn$ explicitly. For more complicated interactions, there is no straightforward way to eliminate $\gmn$ from the bimetric action.

Turning to the equations of motion for the second metric, one finds that variation of the action (\ref{auxeibi}) with respect to $\fmn$ gives,
\beqn\label{feq}
0=\frac{1}{\kappa^2}\sqrt{g}~g^{\mu\nu}-\frac{\lambda}{\kappa^2}\sqrt{f}~\Lambda_f f^{\mu\nu}+\sqrt{f}~T^{\mu\nu}\,,
\eeqn
where we have again used the definition for the stress-energy tensor, $\sqrt{f}\,T^{\mu\nu}=\frac{\partial\big(\sqrt{f}\,\mathcal{L}_\mathrm{matter})}{\partial f_{\mu\nu}}$. This equation is identical to~(\ref{eibieqs2}), showing that the above bimetric action indeed reproduces the complete set of equations for EiBI theory. 

Notice that the action~(\ref{auxeibi}) does not have the ghost-free structure of~(\ref{biact}) and~(\ref{pot}) and, at first sight, these two bimetric theories seem unrelated.
In order to reveal a connection between the two, we will show in the following how both interaction potentials can in fact be derived from the same underlying theory, namely the auxiliary field construction we developed in the previous section.

\subsection{Recovering the EiBI action}

In the following we will work with the auxiliary action (\ref{genauxact}) and focus on the case $\beta_3=0$. Moreover, we remove the Einstein-Hilbert term for $\fmn$ by setting $m_f=0$ and write the Einstein-Hilbert term in Palatini form by introducing an independent connection field $\Gamma^\mu_{\rho\sigma}$.
The action becomes,
\beqn\label{master}
S[g,f,\Phi]&=&m_g^2\int\dd^4x\sqrt{g}\,\big(g^{\mu\nu}\rmn(\Gamma)-2\Lambda_g\big)-~2m^2m_g^2\int\dd^4x\,\big(V_1+V_2\big)\nn\\
&~&~-2m_f^2\int\dd^4x\sqrt{f}~\Lambda_f +\int\dd^4x\sqrt{f}~\mathcal{L}_\mathrm{matter}(f)\,,
\eeqn
with $V_1+V_2$ given in (\ref{potdef}). Also here, the equations of motion for the connection field have the solution $\Gamma^\mu_{\rho\sigma}=\frac{1}{2}g^{\mu\lambda}(\partial_\rho g_{\sigma\lambda}+\partial_\sigma g_{\rho\lambda}-\partial_\lambda g_{\rho\sigma})$, and hence integrating out $\Gamma$ simply results in the Ricci scalar for the metric $\gmn$. 
This action looks similar to the action for generalized dRGT massive gravity with an arbitrary reference metric $\fmn$ that is coupled to matter. However, in the dRGT setup, $\fmn$ is entirely nondynamical, in the sense that one does not evaluate its equations of motion but rather fixes its form by hand. Furthermore, whenever matter is present it couples to the dynamical metric $\gmn$ and not to $\fmn$. Since here we are in the framework of bimetric theory, we do vary the action with respect to $\fmn$ which now, due to the absence of its kinetic term, essentially will act as another auxiliary field.

We proceed by integrating out the auxiliary fields $\Phi^\mu_{~\nu}$. Their equations of motion were already given in (\ref{bteq}) and we found that they gave rise to two branches of solutions,
\beqn\label{brbb2}
{\text{\bf(I)}}~~\Phi^\mu_{~\nu}=\pm\big(\sqrt{g^{-1}f}\,\big)^\mu_{~\nu}\,,\qquad \text{and}\qquad
{\text{\bf(II)}}~~\Tr\Big[ \Phi+\Phi^{-1}g^{-1}f  \Big]=-\frac{2\beta_1}{\beta_2}\,.
\eeqn
After integrating out $\Phi^\mu_{~\nu}$ using the branch~{\bf I} solution, we arrive at the ghost-free bimetric action~(\ref{biact}) with $\beta_3=m_f=0$.
On the other hand, using the branch~{\bf II} solution gives,
\beqn\label{brtei}
S^{(\mathrm{II})}[g,f]&=&m_g^2\int\dd^4x\sqrt{g}\,\big(g^{\mu\nu}\rmn(\Gamma)-2\Lambda\big)+m^2m_g^2\int\dd^4x\sqrt{g}~\Tr\big[g^{-1}f\big]\nn\\
&~&~-2m_f^2\int\dd^4x\sqrt{f}~\Lambda_f +\int\dd^4x\sqrt{f}~\mathcal{L}_\mathrm{matter}(f)\,,
\eeqn
where $\Lambda=\Lambda_g-\frac{\beta_1^2m^2}{2\beta_2}$. This is again a bimetric action but it does not have the ghost-free structure in its interaction potential. However, we now notice that its form is identical to that of the auxiliary action for EiBI theory in~(\ref{auxeibi}).
In particular if we make the identifications,
\beqn
m_g^2=\frac{1}{\kappa}\,,\qquad \Lambda_fm_f^2=\frac{\lambda}{\kappa^2}\,,
\eeqn
and furthermore choose to fix $\Lambda_g$ and $m^2$ according to,
\beqn\label{para}
\Lambda_g=\frac{1}{\kappa}+\frac{\beta_1^2m^2}{2\beta_2}\,,
\qquad m^2=\frac{1}{\kappa}\,,
\eeqn
then the action (\ref{brtei}) precisely matches~(\ref{auxeibi}). This demonstrates that the generalized bimetric action~(\ref{master}) with auxiliary fields includes both the ghost-free bimetric action~(\ref{biact}) with parameters~(\ref{para}) and $\beta_3=m_f=0$ as well as the auxiliary action for EiBI theory~(\ref{eibiactor}). Note that this conclusion remains even when $\beta_1=0$, i.e.~when only $\beta_2$ is nonzero.

\section{Summary $\&$ discussion}\label{discussion}

In this paper we have extended the ghost-free bimetric action of~\cite{Hassan:2011zd} by introducing auxiliary fields into the interaction potential, which allowed us to remove the square-root matrix from the action. Upon integrating out the auxiliary fields it is always possible to recover the ghost-free structure as part of the solution space which we refer to as branch~{\bf I}. 
The existence of a branch~{\bf II} and the possibility to fully integrate out the auxiliary fields on this branch depends on the parameter choices in the interaction potential~(\ref{potdef}):

\begin{itemize}
\item For $\beta_2=0$ branch~{\bf II} does not exist in general and the auxiliary formulation is classically equivalent to ghost-free bimetric theory. 

\item For $\beta_3=0$ (or $\beta_1=0$, depending on in which way the auxiliary fields are introduced, c.f.~the discussion of section \ref{sec:auxamb}) the auxiliary fields can be fully eliminated from the action using the branch~{\bf II} solution. The resulting bimetric action is different from the ghost-free one, but interestingly is capable of exactly reproducing Eddington-inspired Born-Infeld theory if we impose certain relations among the parameters of the theory. This connection still holds if $\beta_1=0$ (or $\beta_3=0$) and is intimately related to having a non-vanishing $\beta_2$.

\item For general parameters, it is not obvious how to obtain a branch~{\bf II} solution for the auxiliary fields which can be used to integrate them out explicitly and obtain a bimetric action. The generic problem is that the branch {\bf II} solutions contain one or more trace combinations of the auxiliary fields, but not the fields themselves. This means that we can at most solve for a specific trace combination, but the action will in general contain other trace combinations as well, so we need a full solution in order to integrate out the fields.
\end{itemize}

As has been pointed out in~\cite{Pani:2013qfa}, gravity models including auxiliary fields face generic phenomenological problems which become visible in the Newtonian limit. These obstacles originate from the fact that the equations of motion in such models can be combined into an Einstein equation but with higher-derivative terms appearing in the matter sector. Generically, such terms lead to an enhanced sensitivity of the metric to abrupt changes in the energy density of matter sources and are therefore tightly constrained by astrophysical observations. The bimetric action for EiBI theory in~(\ref{auxeibi}) is an example of a model in which the aforementioned problems have been shown to arise~\cite{Pani:2012qd}. A way to overcome this problem could be to give full dynamics to the metric $\fmn$ which acts as an auxiliary field in~(\ref{auxeibi}). For instance, one could add an Einstein-Hilbert term $\sqrt{f}\,R(f)$ to the action. Interestingly, this would not spoil the possibility of integrating out $\gmn$ and arrive at the Born-Infeld structure. The resulting action would be of the same form as~(\ref{eibiactor}) but with the extra kinetic term for~$\fmn$. In its bimetric formulation, this type of theory has been under consideration since the work of Isham, Salam and Strathdee in the 70's~\cite{Isham:1971gm}. Its cosmology has been studied e.g.~in~\cite{Banados:2008fj}, where it was shown that the second metric can act as both dark energy and dark matter.

Unfortunately, there is another major drawback to Eddington-inspired Born-Infeld theory. Since its equivalent bimetric formulation does not possess the unique ghost-free structure of~\cite{Hassan:2011zd}, it is suspected to propagate the additional degree of freedom known as the Boulware-Deser ghost~\cite{Boulware:1972zf, Boulware:1973my}. This instability problem does not occur when matter is absent and there is no kinetic term for the second metric because in this case the theory is equivalent to general relativity. The introduction of matter destroys this equivalence, but it is not entirely obvious if the bimetric action~(\ref{auxeibi}) without kinetic term for the metric that couples to matter suffers from the ghost instability or not. Nevertheless, latest when a second Einstein-Hilbert term is included in order to avoid the above astrophysical problems, the extra ghost mode starts propagating. For our auxiliary formulation this implies that the branch of solutions corresponding to EiBI theory does not lead to a consistent theory. This does, however, not imply that the auxiliary field action~(\ref{genauxact}) exhibits a ghost instability around general backgrounds. In fact, we know that this is not the case since the branch~{\bf I} solutions have been shown to be ghost-free. It would therefore be useful to extend the ADM analysis of~\cite{Golovnev:2011aa, Hassan:2012qv} to confirm the status of ghosts in the full auxiliary field formulation.


\vspace{10pt}

\noindent
{\bf Acknowledgments:} The authors would like to thank Pedro Ferreira for helpful discussions. The research of MvS leading to these results has received
funding from the European Research Council under
the European Communitys Seventh Framework Programme
(FP7/2007-2013 Grant Agreement no. 307934).



\end{document}